\documentclass[aps,pra,twocolumn,nofootinbib,longbibliography]{revtex4-2}
\usepackage{bm,amsmath,amssymb,graphicx}
\usepackage[colorlinks,linkcolor=blue,urlcolor=blue]{hyperref}
\newcommand{\Dbra}[1]{\left\langle#1\right|}
\newcommand{\Dket}[1]{\left|#1\right\rangle}

\begin{document}

\title{Quantum entanglement without superposition}

\author{Hans Christian \"Ottinger}
\email[]{hco@mat.ethz.ch}
\homepage[]{www.polyphys.mat.ethz.ch}
\affiliation{ETH Z\"urich, Department of Materials, CH-8093 Z\"urich, Switzerland}

\date{\today}

\begin{abstract}
Superposition states are at the origin of many paradoxes in quantum mechanics. By unraveling the von Neumann equation for density matrices, we develop a superposition-free formulation of quantum mechanics. Stochastic quantum jumps are a key feature of this approach, in blatant contrast with the continuity of the deterministic Schr\"odinger equation. We explain how quantum entanglement arises. Our superposition-free formulation results offers a new perspective on quantum mechanics.
\end{abstract}

\maketitle

\paragraph*{Introduction.}
Many of the famous paradoxes in quantum mechanics begin with the assumption of superpositions of states. The usual discussion of the Einstein-Podolsky-Rosen (EPR) \emph{Gedankenexperiment} \cite{EPR35} in Bohm's spin version is based on the singlet state of a pair of spin-$\frac{1}{2}$ particles \cite{Aspectetal82,Bertlmann90,Aspect02ip}. In Schr\"odinger's cat version, the superposition state is considerably more abstract by invoking complex systems. In the tradition of ``Wigner's friend'' \cite{Wigner61ip}, Frauchiger and Renner \cite{FrauchigerRenner18} assume the applicability of quantum mechanics to even more complex superposition states that include agents who are themselves using quantum theory, and they then reveal inconsistencies in conventional quantum mechanics.

Whereas entanglement is an essential feature of quantum physics, superposition is not. It is the purpose of this letter to propose an alternative, superposition-free implementation of quantum entanglement, thus eliminating a prolific source of paradoxes in conventional quantum mechanics. The mathematical details of the alternative approach have recently been developed in the context of a stochastic simulation technique for reversible quantum dynamics \cite{hco251}. We here elaborate on the implications of that simulation technique for the foundations and interpretation of quantum mechanics.

\paragraph*{The von Neumann equation.}
The stochastic nature of quantum mechanics calls for an appropriate setting to describe randomness. The most natural choice is given by a density matrix $\rho$ on a Hilbert space for a quantum system of interest, by which the average of any observable $A$ can be obtained as a trace, ${\rm tr}(\rho A)$. In the Schr\"odinger picture, the evolution of the density matrix $\rho_t$ describing the time-dependent state of a quantum system is given by the von Neumann equation
\begin{equation}\label{vonNeumanneq}
   i \hbar \frac{d \rho_t}{dt} = [H,\rho_t] ,
\end{equation}
where $\hbar$ is the reduced Planck constant, $H$ is the Hamiltonian of the system, and the square brackets denote the commutator. Except for its greater flexibility in choosing initial conditions, the von Neumann equation is equivalent to the Schr\"odinger equation. Quantum master equations for density matrices have the additional advantage that they are perfectly suited not only for describing reversible dynamics, but also for dissipative quantum systems \cite{BreuerPetru,Weiss}. A class of robust quantum master equations for dissipative systems has been obtained by quantizing the geometric structures behind classical nonequilibrium thermodynamics \cite{hco199,hco221}.

Schr\"odinger's complex wave function $\psi_t$ is less directly associated with the randomness of quantum mechanics than the density matrix $\rho_t$. One needs the additional rule that the squared modulus of the wave function, $|\psi_t|^2$, should be interpreted as a probability density, whereas Eq.~(\ref{vonNeumanneq}) is linear in the primary probabilistic quantity of interest, $\rho_t$. When one solves the continuous Schr\"odinger equation for an interacting system, superposition of states seems to occur inevitably. For quantum master equations, an alternative option is available, which we discuss next.

\paragraph*{Two-process unravelings.}
The general strategy of so-called unravelings is to introduce stochastic processes in Hilbert space such that a density matrix evolving according to the von Neumann equation (\ref{vonNeumanneq}) can be extracted in terms of suitable averages. This idea of passing from probabilistic concepts to equivalent stochastic processes was originally motivated by computer simulations for dissipative quantum systems \cite{BreuerPetru} but, as we show here, it is useful also for reversible systems and conceptual clarifications. Unravelings can even offer an alternative interpretation of quantum mechanics. The passage from probabilities to stochastic objects is particularly relevant to ontology \cite{hco243}. We here focus on the opportunity to obtain entanglement without superposition, with the underlying goal of avoiding paradoxes in quantum mechanics.

For dissipative systems, unravelings employ continuous evolution for reversible dynamics and a combination of continuous evolution and stochastic jump processes for dissipative dynamics \cite{BreuerPetru}. In the context of dissipative quantum field theory \cite{hcoqft}, it has been proposed to treat also reversible interactions by jumps. This idea has been elaborated in detail for simulating purely reversible quantum dynamics in \cite{hco251}.

The interpretation of interactions as jumps requires a splitting of the full Hamiltonian into free and interacting parts, $H = H^{\rm free} + H^{\rm int}$, and a distinguished basis $\cal B$ of orthonormal eigenstates of the free Hamiltonian $H^{\rm free}$. For unravelings in terms of two stochastic processes $\Dket{\phi}_t$ and $\Dket{\psi}_t$ in Hilbert space, one wishes to reproduce the density matrix $\rho_t$ by the following expectation,
\begin{equation}\label{unravel}
   \rho_t = E\big( \Dket{\phi}_t \Dbra{\psi}_t \big) ,
\end{equation}
where we have used Dirac's bra-ket notation. In this setting we can impose the strict superselection rule that the states $\Dket{\phi}_t$ and $\Dket{\psi}_t$ can only be multiples of basis vectors from $\cal B$. Reasons for believing in the existence of such a distinguished basis will be offered below.

The strict superselection rule has the pleasing side-effect of reducing the enormous number of possibilities for constructing unravelings. It naturally guides us to the following construction of piecewise continuous trajectories with interspersed jumps occurring at a rate $r$:
\\[2mm]
\underline{Continuous evolution} results from solving separate Schr\"odinger equations for $\Dket{\phi}_t$ and $\Dket{\psi}_t$ for the free Hamiltonian $H^{\rm free}$. The basis vectors remain unchanged. Their prefactors are modified by exponential phase factors that are determined by the corresponding eigenvalues of $H^{\rm free}$.
\\[1mm]
\underline{Quantum jumps} occurring at rate $r$ reproduce the effect of the interaction Hamiltonian $H^{\rm int}$. With probability $1/2$ a jump occurs for one of the vectors $\Dket{\phi}_t$ or $\Dket{\psi}_t$. The transition from an initial base vector $b \in \cal B$ to a particular final base vector $b' \in \cal B$ then occurs with a probability proportional to $|\Dbra{b'} H^{\rm int} \Dket{b}|$. The prefactor inherited by $b'$ from $b$ needs to be modified by a proper normalization factor.
\\[2mm]
This level of detail in introducing unravelings should be sufficient for the conceptual issues discussed in this letter. For more details, curious readers could look at the actual implementation of unravelings in computer simulations of reversible quantum dynamics \cite{hco251}.

The continuous evolution shows the importance of allowing for scalar multiplication. The jumps eliminate the need for any addition of states.

\paragraph*{Triplet unravelings.}
Two-process unravelings might work perfectly well on the analog quantum computer called ``Nature,'' which seems to possess unlimited computational resources. For man-made computers, however, efficiency is a major issue, in particular in view of the infamous sign problem plaguing many quantum simulations \cite{TroyerWiese05}.

Actual simulations of reversible quantum dynamics based on superposition-free unravelings have been performed in \cite{hco251}. For reasons of efficiency, these simulations are performed in Laplace-transformed time. Importance sampling and approximations have been introduced to alleviate the sign problem. In practice, the sign problem can be addressed more efficiently by switching from two-process unravelings to triplet unravelings. In the triplets $(c_t,\Dket{\phi}_t,\Dket{\psi}_t)$, the states $\Dket{\phi}_t$ and $\Dket{\psi}_t$ are elements of the discrete set of distinguished basis vectors without any prefactors, and the products of previously allowed prefactors are combined into the complex numbers $c_t$. Ensembles can then be compressed efficiently by adding up all the factors $c_t$ for a given pair of base vectors.

\paragraph*{Initial conditions.}
We have focussed on reproducing the dynamics of the von Neumann equation by unravelings. How about initial conditions?

Let us consider a two-dimensional Hilbert space with orthonormal basis ${\cal B} = \{b_1,b_2\}$ for illustration. Let us assume that the initial density matrix $\rho_0$ is associated with the superposition state $(b_1+b_2)/\sqrt{2}$. Can such a density matrix for a pure state be unraveled without violating the strict superposition principle? To obtain a positive answer, assume that the random variable $\Dket{\phi}_0$ takes the values $\sqrt{2} b_1$ and $\sqrt{2} b_2$ with probability $1/2$, and assume the same independently for $\Dket{\psi}_0$; then, the desired density matrix $\rho_0$ of rank $1$ is recovered from Eq.~(\ref{unravel}). In conclusion, we have shown how to  represent entangled initial conditions and their time evolution without any superpositions.

\paragraph*{Limitations of quantum mechanics.}
In conventional quantum mechanics one is free to do basis changes, which typically involve superpositions. Therefore, we need to justify our assumption of the existence of a distinguished basis $\cal B$ in superposition-free quantum mechanics. To this end one needs to recognize the limitations and the approximate nature of quantum mechanics and to look at its deeper origin.

Dirac's fascinating quantization recipe formally suggests that we can quantize any classical Hamiltonian system, preferably in canonical coordinates, by replacing classical Poisson brackets by quantum commutators. However, the quantization of say planetary motion can hardly be particularly meaningful. A restriction of quantum theory to fundamental particles, which would be the other extreme, is unpractical because we can never be sure that we know any fundamental particles. For example, nobody hesitates to apply quantum mechanics to protons.

A nice textbook example for the application of quantum mechanics is the hydrogen atom, which leads to the famous prediction for the line spectrum of atomic hydrogen. Note that the Hamiltonian for this problem contains the $1/r$ Coulomb potential of classical electrostatics. In that sense, quantum mechanics has a semi-classical character. Particularly questionable is the occurrence of electron-proton interactions at a distance.

In a more rigorous treatment of the hydrogen atom, the Coulomb potential should result from an exchange of photons between the proton, or the quarks in a proton, and an electron. We are then in the domain of quantum field theory, allowing for the creation and annihilation of photons. Quantum mechanics can only be an approximate theory that arises in the low-energy limit of quantum field theory, when particle creation and annihilation are suppressed. Quantum mechanics should emerge from the quantum field theory of fundamental particles and interactions after making approximations.

In quantum field theory, Fock space comes with a natural basis $\cal F$ \cite{Fock32,Teller,hcoqft}. The Fock basis vectors describe states with well-defined (free) particle content. The strict superselection rule implies that a system can never be in a superposition state of different particle contents. In making approximations for passing from quantum field theory to quantum mechanics one should make sure that a distinguished basis $\cal B$ emerges from the natural Fock basis $\cal F$. For example, the treatment of a proton as a fundamental particle on the atomic length scale (or low energies) in quantum mechanics relies on color confinement.

\paragraph*{EPR experiment.}
As an example of getting guidance from quantum field theory in quantum mechanics, we revisit the EPR experiment in the version for photons \cite{Aspectetal82}. Pairs of photons moving with the same circular polarization in opposite directions can be created in the decay of properly excited calcium atoms. Whereas the actual decay occurs through an intermediate state within some $5\,{\rm ns}$, we here use the idealization of a single collision in which a calcium excitation is annihilated and two photons are created.  

We start the unraveling with a calcium excitation, from which a photon pair is created eventually, rather than from the usually assumed superposition state. By evolving the initial state via the two-process unraveling, a pair of left- or right-handed photons can be created in the $\Dket{\phi}$ and $\Dket{\psi}$ states. If re-excitations of calcium atoms can be neglected, this situation corresponds exactly to the example considered in the above discussion of initial conditions.

\paragraph*{Measurements.}
In the context of linear quantum master equations, a general theory of measurable multitime correlation functions has been developed in \cite{BreuerPetru}. For the semilinear quantum master equations of thermodynamic origin, which are linear for scalar multiplication but, in general, nonlinear for addition, this theory of measurable correlations has been generalized in \cite{hcoqft}. Note that this kind of semilinearity goes nicely together with the strict superposition rule. A detailed discussion of measurements is beyond te scope of this letter.

The idea of eigenstates of an observable $A$ is generalized to the set of density matrices $\rho$ that commute with $A$. Such density matrices characterize the superposition-free mixtures of eigenstates. If, for such a density matrix, $A \rho = a \rho$ holds with $a \in \mathbb R$, then $a$ is an eigenvalue.

One can ask the question whether density matrices can be measured. This question can, for example, be addressed by quantum-state tomography \cite{Rau10,Rau}.

\paragraph*{Conclusions.}
We have shown how to eliminate superposition from quantum mechanics by imposing the strongest possible superselection rule. In our approach to reversible quantum dynamics based on two-process unravelings, we need to solve the two free Schr\"odinger equations, $i \hbar \partial \Dket{\phi}_t / \partial t = H^{\rm free} \Dket{\phi}_t$ and $i \hbar \partial \Dket{\psi}_t / \partial t = H^{\rm free} \Dket{\psi}_t$ (with trivial solutions), interrupted by stochastic quantum jumps between multiples of base vectors due to interactions. The density matrix is then given by Eq.~(\ref{unravel}). In such a stochastic unraveling, the interrupted Schr\"odinger dynamics is more elementary than in the GRW approach \cite{GhirardiRimWeb86}.

Unravelings provide an opportunity of a new interpretation of quantum mechanics. In the context of dissipative quantum field theory, this opportunity has been explored in \cite{hcoqft,hco243}.

In the proposed superposition-free approach, entanglement arises from the interplay of two stochastic processes in Hilbert space. Complete knowledge of a quantum state requires two state vectors. Tow ``semi-worlds'' play together to characterize the quantum state of the "full world,'' including entanglements.


\begin{thebibliography}{19}%
\makeatletter
\providecommand \@ifxundefined [1]{%
 \@ifx{#1\undefined}
}%
\providecommand \@ifnum [1]{%
 \ifnum #1\expandafter \@firstoftwo
 \else \expandafter \@secondoftwo
 \fi
}%
\providecommand \@ifx [1]{%
 \ifx #1\expandafter \@firstoftwo
 \else \expandafter \@secondoftwo
 \fi
}%
\providecommand \natexlab [1]{#1}%
\providecommand \enquote  [1]{``#1''}%
\providecommand \bibnamefont  [1]{#1}%
\providecommand \bibfnamefont [1]{#1}%
\providecommand \citenamefont [1]{#1}%
\providecommand \href@noop [0]{\@secondoftwo}%
\providecommand \href [0]{\begingroup \@sanitize@url \@href}%
\providecommand \@href[1]{\@@startlink{#1}\@@href}%
\providecommand \@@href[1]{\endgroup#1\@@endlink}%
\providecommand \@sanitize@url [0]{\catcode `\\12\catcode `\$12\catcode
  `\&12\catcode `\#12\catcode `\^12\catcode `\_12\catcode `\%12\relax}%
\providecommand \@@startlink[1]{}%
\providecommand \@@endlink[0]{}%
\providecommand \url  [0]{\begingroup\@sanitize@url \@url }%
\providecommand \@url [1]{\endgroup\@href {#1}{\urlprefix }}%
\providecommand \urlprefix  [0]{URL }%
\providecommand \Eprint [0]{\href }%
\providecommand \doibase [0]{https://doi.org/}%
\providecommand \selectlanguage [0]{\@gobble}%
\providecommand \bibinfo  [0]{\@secondoftwo}%
\providecommand \bibfield  [0]{\@secondoftwo}%
\providecommand \translation [1]{[#1]}%
\providecommand \BibitemOpen [0]{}%
\providecommand \bibitemStop [0]{}%
\providecommand \bibitemNoStop [0]{.\EOS\space}%
\providecommand \EOS [0]{\spacefactor3000\relax}%
\providecommand \BibitemShut  [1]{\csname bibitem#1\endcsname}%
\let\auto@bib@innerbib\@empty
\bibitem [{\citenamefont {Einstein}\ \emph {et~al.}(1935)\citenamefont
  {Einstein}, \citenamefont {Podolsky},\ and\ \citenamefont {Rosen}}]{EPR35}%
  \BibitemOpen
  \bibfield  {author} {\bibinfo {author} {\bibfnamefont {A.}~\bibnamefont
  {Einstein}}, \bibinfo {author} {\bibfnamefont {B.}~\bibnamefont {Podolsky}},\
  and\ \bibinfo {author} {\bibfnamefont {N.}~\bibnamefont {Rosen}},\ }\bibfield
   {title} {\bibinfo {title} {Can quantum-mechanical description of physical
  reality be considered complete?},\ }\href
  {https://doi.org/10.1103/PhysRev.47.777} {\bibfield  {journal} {\bibinfo
  {journal} {Phys.\ Rev.}\ }\textbf {\bibinfo {volume} {47}},\ \bibinfo {pages}
  {777} (\bibinfo {year} {1935})}\BibitemShut {NoStop}%
\bibitem [{\citenamefont {Aspect}\ \emph {et~al.}(1982)\citenamefont {Aspect},
  \citenamefont {Grangier},\ and\ \citenamefont {Roger}}]{Aspectetal82}%
  \BibitemOpen
  \bibfield  {author} {\bibinfo {author} {\bibfnamefont {A.}~\bibnamefont
  {Aspect}}, \bibinfo {author} {\bibfnamefont {P.}~\bibnamefont {Grangier}},\
  and\ \bibinfo {author} {\bibfnamefont {G.}~\bibnamefont {Roger}},\ }\bibfield
   {title} {\bibinfo {title} {Experimental realization of
  {E}instein-{P}odolsky-{R}osen-{B}ohm {G}edankenexperiment: {A} new violation
  of {B}ell's inequalities},\ }\href
  {https://doi.org/10.1103/PhysRevLett.49.91} {\bibfield  {journal} {\bibinfo
  {journal} {Phys.\ Rev.\ Lett.}\ }\textbf {\bibinfo {volume} {49}},\ \bibinfo
  {pages} {91} (\bibinfo {year} {1982})}\BibitemShut {NoStop}%
\bibitem [{\citenamefont {Bertlmann}(1990)}]{Bertlmann90}%
  \BibitemOpen
  \bibfield  {author} {\bibinfo {author} {\bibfnamefont {R.~A.}\ \bibnamefont
  {Bertlmann}},\ }\bibfield  {title} {\bibinfo {title} {Bell's theorem and the
  nature of reality},\ }\href {https://doi.org/10.1007/BF01889465} {\bibfield
  {journal} {\bibinfo  {journal} {Found.\ Phys.}\ }\textbf {\bibinfo {volume}
  {20}},\ \bibinfo {pages} {1191} (\bibinfo {year} {1990})}\BibitemShut
  {NoStop}%
\bibitem [{\citenamefont {Aspect}(2002)}]{Aspect02ip}%
  \BibitemOpen
  \bibfield  {author} {\bibinfo {author} {\bibfnamefont {A.}~\bibnamefont
  {Aspect}},\ }\bibfield  {title} {\bibinfo {title} {Bell’s theorem: {T}he
  naive view of an experimentalist},\ }in\ \emph {\bibinfo {booktitle}
  {\href {https://doi.org/10.1007/978-3-662-05032-3} {Quantum (Un)speakables:}
  {F}rom {B}ell to Quantum Information}},\ \bibinfo
  {editor} {edited by\ \bibinfo {editor} {\bibfnamefont {R.~A.}\ \bibnamefont
  {Bertlmann}}\ and\ \bibinfo {editor} {\bibfnamefont {A.}~\bibnamefont
  {Zeilinger}}}\ (\bibinfo  {publisher} {Springer},\ \bibinfo {address}
  {Berlin},\ \bibinfo {year} {2002})\ pp.\ \bibinfo {pages}
  {119--153}\BibitemShut {NoStop}%
\bibitem [{\citenamefont {Wigner}(1961)}]{Wigner61ip}%
  \BibitemOpen
  \bibfield  {author} {\bibinfo {author} {\bibfnamefont {E.}~\bibnamefont
  {Wigner}},\ }\bibfield  {title} {\bibinfo {title} {Remarks on the mind-body
  question},\ }in\ \emph
  {\bibinfo {booktitle} {\href {https://doi.org/10.1007/978-3-642-78374-6_20} {The Scientist Speculates:} An Anthology of Partly-Baked
  Ideas}},\ \bibinfo {editor} {edited by\ \bibinfo {editor} {\bibfnamefont
  {I.~J.}\ \bibnamefont {Good}}}\ (\bibinfo  {publisher} {Heinemann},\ \bibinfo
  {address} {London},\ \bibinfo {year} {1961})\ pp.\ \bibinfo {pages}
  {284--302}\BibitemShut {NoStop}%
\bibitem [{\citenamefont {Frauchiger}\ and\ \citenamefont
  {Renner}(2018)}]{FrauchigerRenner18}%
  \BibitemOpen
  \bibfield  {author} {\bibinfo {author} {\bibfnamefont {D.}~\bibnamefont
  {Frauchiger}}\ and\ \bibinfo {author} {\bibfnamefont {R.}~\bibnamefont
  {Renner}},\ }\bibfield  {title} {\bibinfo {title} {Quantum theory cannot
  consistently describe the use of itself},\ }\href
  {https://doi.org/10.1038/s41467-018-05739-8} {\bibfield  {journal} {\bibinfo
  {journal} {Nature Communications}\ }\textbf {\bibinfo {volume} {9}},\
  \bibinfo {pages} {3711} (\bibinfo {year} {2018})}\BibitemShut {NoStop}%
\bibitem [{\citenamefont {Chessex}\ \emph {et~al.}(2022)\citenamefont
  {Chessex}, \citenamefont {Borrelli},\ and\ \citenamefont
  {{\"O}ttinger}}]{hco251}%
  \BibitemOpen
  \bibfield  {author} {\bibinfo {author} {\bibfnamefont {R.}~\bibnamefont
  {Chessex}}, \bibinfo {author} {\bibfnamefont {M.}~\bibnamefont {Borrelli}},\
  and\ \bibinfo {author} {\bibfnamefont {H.~C.}\ \bibnamefont {{\"O}ttinger}},\
  }\bibfield  {title} {\bibinfo {title} {Dynamical triplet unraveling: {A}
  quantum {M}onte {C}arlo algorithm for reversible dynamics},\ }\href
  {https://doi.org/10.1103/PhysRevA.106.022222} {\bibfield  {journal} {\bibinfo
   {journal} {Phys.\ Rev.\ A}\ }\textbf {\bibinfo {volume} {106}},\ \bibinfo
  {pages} {022222} (\bibinfo {year} {2022})}\BibitemShut {NoStop}%
\bibitem [{\citenamefont {Breuer}\ and\ \citenamefont
  {Petruccione}(2002)}]{BreuerPetru}%
  \BibitemOpen
  \bibfield  {author} {\bibinfo {author} {\bibfnamefont {H.-P.}\ \bibnamefont
  {Breuer}}\ and\ \bibinfo {author} {\bibfnamefont {F.}~\bibnamefont
  {Petruccione}},\ }\href
  {https://doi.org/10.1093/acprof:oso/9780199213900.001.0001} {\emph {\bibinfo
  {title} {The Theory of Open Quantum Systems}}}\ (\bibinfo  {publisher}
  {Oxford University Press},\ \bibinfo {address} {Oxford},\ \bibinfo {year}
  {2002})\BibitemShut {NoStop}%
\bibitem [{\citenamefont {Weiss}(2008)}]{Weiss}%
  \BibitemOpen
  \bibfield  {author} {\bibinfo {author} {\bibfnamefont {U.}~\bibnamefont
  {Weiss}},\ }\href {https://doi.org/10.1142/6738} {\emph {\bibinfo {title}
  {Quantum Dissipative Systems}}},\ \bibinfo {edition} {3rd}\ ed.,\ Series in
  Modern Condensed Matter Physics, Volume~13\ (\bibinfo  {publisher} {World
  Scientific},\ \bibinfo {address} {Singapore},\ \bibinfo {year}
  {2008})\BibitemShut {NoStop}%
\bibitem [{\citenamefont {{\"O}ttinger}(2011)}]{hco199}%
  \BibitemOpen
  \bibfield  {author} {\bibinfo {author} {\bibfnamefont {H.~C.}\ \bibnamefont
  {{\"O}ttinger}},\ }\bibfield  {title} {\bibinfo {title} {The geometry and
  thermodynamics of dissipative quantum systems},\ }\href
  {https://doi.org/10.1209/0295-5075/94/10006} {\bibfield  {journal} {\bibinfo
  {journal} {Europhys.\ Lett.}\ }\textbf {\bibinfo {volume} {94}},\ \bibinfo
  {pages} {10006} (\bibinfo {year} {2011})}\BibitemShut {NoStop}%
\bibitem [{\citenamefont {Taj}\ and\ \citenamefont
  {{\"O}ttinger}(2015)}]{hco221}%
  \BibitemOpen
  \bibfield  {author} {\bibinfo {author} {\bibfnamefont {D.}~\bibnamefont
  {Taj}}\ and\ \bibinfo {author} {\bibfnamefont {H.~C.}\ \bibnamefont
  {{\"O}ttinger}},\ }\bibfield  {title} {\bibinfo {title} {Natural approach to
  quantum dissipation},\ }\href {https://doi.org/10.1103/PhysRevA.92.062128}
  {\bibfield  {journal} {\bibinfo  {journal} {Phys.\ Rev.\ A}\ }\textbf
  {\bibinfo {volume} {92}},\ \bibinfo {pages} {062128} (\bibinfo {year}
  {2015})}\BibitemShut {NoStop}%
\bibitem [{\citenamefont {Oldofredi}\ and\ \citenamefont
  {{\"O}ttinger}(2021)}]{hco243}%
  \BibitemOpen
  \bibfield  {author} {\bibinfo {author} {\bibfnamefont {A.}~\bibnamefont
  {Oldofredi}}\ and\ \bibinfo {author} {\bibfnamefont {H.~C.}\ \bibnamefont
  {{\"O}ttinger}},\ }\bibfield  {title} {\bibinfo {title} {The dissipative
  approach to quantum field theory: Conceptual foundations and ontological
  implications},\ }\href {https://doi.org/10.1007/s13194-020-00330-9}
  {\bibfield  {journal} {\bibinfo  {journal} {Euro.\ Jnl.\ Phil.\ Sci.}\
  }\textbf {\bibinfo {volume} {11}},\ \bibinfo {pages} {18} (\bibinfo {year}
  {2021})}\BibitemShut {NoStop}%
\bibitem [{\citenamefont {{\"O}ttinger}(2017)}]{hcoqft}%
  \BibitemOpen
  \bibfield  {author} {\bibinfo {author} {\bibfnamefont {H.~C.}\ \bibnamefont
  {{\"O}ttinger}},\ }\emph
  {\bibinfo {title} {\href {https://doi.org/10.1017/9781108227667} {A Philosophical Approach} to Quantum Field Theory}}\
  (\bibinfo  {publisher} {Cambridge University Press},\ \bibinfo {address}
  {Cambridge},\ \bibinfo {year} {2017})\BibitemShut {NoStop}%
\bibitem [{\citenamefont {Troyer}\ and\ \citenamefont
  {Wiese}(2005)}]{TroyerWiese05}%
  \BibitemOpen
  \bibfield  {author} {\bibinfo {author} {\bibfnamefont {M.}~\bibnamefont
  {Troyer}}\ and\ \bibinfo {author} {\bibfnamefont {U.-J.}\ \bibnamefont
  {Wiese}},\ }\bibfield  {title} {\bibinfo {title} {Computational complexity
  and fundamental limitations to fermionic quantum {M}onte {C}arlo
  simulations},\ }\href {https://doi.org/10.1103/PhysRevLett.94.170201}
  {\bibfield  {journal} {\bibinfo  {journal} {Phys.\ Rev.\ Lett.}\ }\textbf
  {\bibinfo {volume} {94}},\ \bibinfo {pages} {170201} (\bibinfo {year}
  {2005})}\BibitemShut {NoStop}%
\bibitem [{\citenamefont {Fock}(1932)}]{Fock32}%
  \BibitemOpen
  \bibfield  {author} {\bibinfo {author} {\bibfnamefont {V.}~\bibnamefont
  {Fock}},\ }\bibfield  {title} {\bibinfo {title} {{K}onfigurationsraum und
  zweite {Q}uantelung},\ }\href {https://doi.org/10.1007/bf01344458} {\bibfield
   {journal} {\bibinfo  {journal} {Z.~Physik}\ }\textbf {\bibinfo {volume}
  {75}},\ \bibinfo {pages} {622} (\bibinfo {year} {1932})}\BibitemShut
  {NoStop}%
\bibitem [{\citenamefont {Teller}(1995)}]{Teller}%
  \BibitemOpen
  \bibfield  {author} {\bibinfo {author} {\bibfnamefont {P.}~\bibnamefont
  {Teller}},\ }\emph {\bibinfo
  {title} {\href {https://doi.org/10.1515/9780691216294} {An Interpretive Introduction} to Quantum Field Theory}}\ (\bibinfo
  {publisher} {Princeton University Press},\ \bibinfo {address} {Princeton},\
  \bibinfo {year} {1995})\BibitemShut {NoStop}%
\bibitem [{\citenamefont {Rau}(2010)}]{Rau10}%
  \BibitemOpen
  \bibfield  {author} {\bibinfo {author} {\bibfnamefont {J.}~\bibnamefont
  {Rau}},\ }\bibfield  {title} {\bibinfo {title} {Evidence procedure for
  efficient quantum-state tomography},\ }\href
  {https://doi.org/10.1103/PhysRevA.82.012104} {\bibfield  {journal} {\bibinfo
  {journal} {Phys.\ Rev.\ A}\ }\textbf {\bibinfo {volume} {82}},\ \bibinfo
  {pages} {012104} (\bibinfo {year} {2010})}\BibitemShut {NoStop}%
\bibitem [{\citenamefont {Rau}(2021)}]{Rau}%
  \BibitemOpen
  \bibfield  {author} {\bibinfo {author} {\bibfnamefont {J.}~\bibnamefont
  {Rau}},\ }\emph
  {\bibinfo {title} {\href {https://doi.org/10.1093/oso/9780192896308.003.0002} {Quantum Theory:} {A}n Information Processing Approach}}\
  (\bibinfo  {publisher} {Oxford University Press},\ \bibinfo {address}
  {Oxford},\ \bibinfo {year} {2021})\BibitemShut {NoStop}%
\bibitem [{\citenamefont {Ghirardi}\ \emph {et~al.}(1986)\citenamefont
  {Ghirardi}, \citenamefont {Rimini},\ and\ \citenamefont
  {Weber}}]{GhirardiRimWeb86}%
  \BibitemOpen
  \bibfield  {author} {\bibinfo {author} {\bibfnamefont {G.~C.}\ \bibnamefont
  {Ghirardi}}, \bibinfo {author} {\bibfnamefont {A.}~\bibnamefont {Rimini}},\
  and\ \bibinfo {author} {\bibfnamefont {T.}~\bibnamefont {Weber}},\ }\bibfield
   {title} {\bibinfo {title} {Unified dynamics for microscopic and macroscopic
  systems},\ }\href {https://doi.org/10.1103/PhysRevD.34.470} {\bibfield
  {journal} {\bibinfo  {journal} {Phys.\ Rev.\ D}\ }\textbf {\bibinfo {volume}
  {34}},\ \bibinfo {pages} {470} (\bibinfo {year} {1986})}\BibitemShut
  {NoStop}%
\end{thebibliography}

%

\end{document}